\documentclass[pra]{revtex4}
\usepackage{amssymb}
\usepackage{amsmath}

\newcommand{\vc}[1]{\boldsymbol{#1}}

\begin{document}

\title{Pure Gauge Spin-Orbit Couplings}

\author{M. S. Shikakhwa}
\affiliation{Physics Group, Middle East Technical University Northern Cyprus Campus,\\
Kalkanl\i, G\"{u}zelyurt, via Mersin 10, Turkey}

\begin{abstract}
Planar systems with a general linear spin-orbit interaction (SOI) that can be cast in the form of  a non-Abelian pure
gauge field are investigated using the language of non-Abelian gauge field theory. A special class of these fields that, though a $2\times2$ matrix, are Abelian are seen to emerge and their general form is given. It is shown that the unitary transformation that gauges away these fields induces at the same time a rotation on the wavefunction about a fixed axis but with a space-dependent angle, both of which being characteristics of the SOI involved.The experimentally important case of  equal-strength Rashba and Dresselhaus SOI (R+D SOI) is shown to fall within this special class of Abelian gauge fields, and  the phenomenon of Persistent Spin Helix (PSH) that emerges in the presence of  this latter SOI in a plane is shown to fit naturally within the general formalism developed. The general formalism is also extended to the case of a particle confined to a ring. It is shown that the Hamiltonian on a ring in the presence of equal-strength R+D SOI is unitarily equivalent to that of a particle subject to only a spin-independent but $\theta$-dependent potential with the unitary transformation relating the two being again the space-dependent rotation operator characteristic of R+D SOI.
\end{abstract}

\maketitle

\section{Introduction}
Planar spin one-half systems with spin-orbit interaction (SOI) have been receiving much attention recently due to their relevance to the important applied field of spintronics \cite{spintronics}. In these systems, the SOI acts as a momentum-dependent  effective magnetic field that gives rise to spin precession, therefore,  manipulation of the spin can be achieved even in the absence of a magnetic field by tuning the electric field that generates the SOI.  So far, two major types of SOC mechanisms have been receiving the most interest; the one resulting from the structure inversion asymmetry, or the Rashba SOI (R) \cite{rashba} , and the one  due to the bulk inversion asymmetry  known as Dresselhaus SOI (D)\cite{dresselhaus}. Systems where both these SOI are present, especially when they are of equal strength, have special importance (see below).
A useful approach to systems with SOI is the non-Abelian gauge field formalism of SOI \cite{frohlich93,Jin.et.al-JPA,Dartora08,Mineev and Volovik} (see also the review \cite{Berche review 1}) . Here, one expresses the SOI interaction as a coupling of the momentum of the particle to a synthetic non-Abelian gauge field. The advantage of this approach is that it provides an elegant and general approach to the problem where one can carry over  the well-established machinery of classical gauge field theory to the treatment and apply  ideas like gauge transformations, gauge-covariance ...etc. One has to keep in mind here, however, that unlike the case in particle physics, the gauge fields in these models, being proportional to the electric fields, are physical. Therefore, a gauge transformation corresponds to a transformation of one physical configuration of these fields to another. With this transformation representing a unitary transformation on the Hamiltonian as we will show below, the corresponding Hamiltonians are unitarily-equivalent and thus iso-spectral.\\
An important symmetry was recently discovered in planar SOI systems where equal-strength R and D SOI's (R+D SOI) are present \cite{exact-su(2)06}. In such systems, the spin is rotated about a fixed direction but with position-dependent angle  such that the projection of the spin of the particle along this direction is constant.  This results in a long-living spin polarization. This phenomenon that became known as Persistent Spin helix (PSH) is important in spintronics applications  and was recently observed experimentally \cite{exp psh 09, exp psh 12}. A gauge field treatment specific only to  R+D SOI's was provided in reference \cite{chen&chang08}, where it was seen that the gauge field in this case is merely a pure gauge, and the PSH was seen to emerge upon gauging it away . No general treatment valid for any linear SOI is present in the literature, however. Moreover, spin one-half systems confined to a ring with R+D SOI were considered in\cite{sheng&chang06}, but without the use of the gauge-field approach. A gauge field treatment with only one or the other of R and D SOI's are present  at a time on a ring was carried out in \cite{Berche review 2}. The present article provides a general gauge field-based  treatment valid for any linear planar SOI that can be expressed as a pure gauge, both for a planar system as well as for a particle confined to a ring. The importance of this general treatment for a planar system is that it makes manifest the fact that the Hamiltonian in the presence of a linear SOI is not, in general, gauge-covariant as far as the non-Abelian gauge field representing the SOI is concerned. Therefore, the use of gauge transformation as done in the literature should be made with caution. We demonstrate the intricacy of this point on general grounds and show how to handle it. This fact becomes important when one considers the case of a ring with pure gauge SOI. Moreover, we address the issue of the conservation of the Coulomb gauge condition satisfied by the non-Abelian gauge field upon gauging away the field, especially in the ring case.\\
In section II, we introduce the Hamiltonian of a spin one-half particle in a general linear SOI that can be expressed as a pure gauge. We, noting the absence of gauge-covariance of this Hamiltonian, discuss how we can still utilize the notion of gauge transformations to gauge away the gauge field from the Hamiltonian. In section III, we consider the special class of pure gauge fields that, though a $2\times 2$ matrix, are Abelian and obtain their general expression demonstrating that the uniary transformation that gauges them away  represents rotation about a fixed axis but with a space-dependent angle. We note that the case of equal strength R+D SOI is merely a special case of this class and thus fit the phenomenon of PSH within our general formalism. In section IV, we  place our Hamiltonian on a ring and show how to consistently gauge away the pure gauge field, demonstrating explicitly how the Coulomb gauge condition is preserved in the process. When applying our formalism again to the case of equal-strength R+D SOI we reproduce the results of reference \cite{sheng&chang06} obtained without using the gauge field formalism. We sum up our results in section V.

\section{Planar system with pure gauge SOI}
The spin-orbit interaction (SOI) emerges upon considering the
$O\left(\frac{1}{c^{2}}\right)$ expansion of the Dirac Hamiltonian
for a spin one-half particle in  a scalar potential
$V\left(\vc{x}\right)$:
\begin{equation}\label{}
H=\frac{\vc{p}^{2}}{2m}+V\left(\vc{x}\right)+\frac{e\hbar}{4m^2c^{2}}\vc{\sigma}\cdot\left(\vc{E}\wedge\vc{p}\right)
   =\frac{\vc{p}^{2}}{2m}+V\left(\vc{x}\right)+\frac{e\hbar}{4m^2c^{2}}\vc{p}\cdot\left(\vc{\sigma}\wedge\vc{E}\right)
\end{equation}
where $\vc{\nabla}\wedge\vc{E}=0$ was assumed. The SOI can be expressed  in the form of  a minimal coupling of an $SU(2)$ field to the momentum operator
by defining the $SU(2)$ gauge field $W_{i}^{a}$ by \cite{shikakhwa12}
\begin{equation}
  -gW_{i}^{a}  \equiv  \frac{e\hbar}{4mc^{2}}\epsilon_{iaj}E_{j} ~~.
  \label{eq:W_as_a_function_of_E_field}
\end{equation}
where $i,a= 1...3$ and $\sigma$'s are the the Pauli matrices. Using $\vc{W}$, the Hamiltonian can be expressed as ( the scalar potential $ V\left(\vc{x}\right)$ is dropped now on) :
\begin{equation}\label{standard H}
  H=\frac{\vc{p}^{2}}{2m}-\frac{g}{m}\vc{p}\cdot\vc{W}=\frac{\vc{p}^{2}}{2m}-\frac{g}{m}\vc{W}\cdot\vc{p}\quad.
\end{equation}
where we have noted that $\vc{\nabla}\wedge\vc{E}=0$ implies $\vc{\nabla}\cdot \vc{W}=0$. Completing the square,  we can put this into the form
\begin{equation}
  H=H(\vc{W})=\frac{\left(\vc{p}-g\vc{W}\right)^{2}}{2m}-\frac{g^{2}}{2m}\vc{W}\cdot\vc{W}
  \label{HNGC}
\end{equation}
where $W_i=W_i^a\sigma_a$ is the $i$th component of the field
$\vc{W}$. The above is the Hamiltonian of a spin one-half particle
coupled to the $SU(2)$ gauge field $\vc{W}$. The spin-independent term
quadratic in the gauge field, i.e.,$-\frac{g^{2}}{2m}\vc{W}\cdot\vc{W}=-\frac{g^{2}}{2m}W_i^aW_i^a=-\frac{e^2\hbar^2}{16m^2c^{2}}E^2$,  evidently breaks the $SU(2)$ gauge symmetry. Some works \cite{Jin.et.al-JPA,Tokatly08} in the literature either drop or absorb this term into the scalar potential thus ending up with a gauge-covariant Hamiltonian:
\begin{equation}\label{gauge covariant H }
 H_{GC}=H_{GC}(\vc{W})=\frac{\left(\vc{p}-g\vc{W}\right)^{2}}{2m}
\end{equation}
This approach  is controversial \cite{medina-epl08,shikakhwa12}, and so we will in this work focus on the non  gauge-covariant  case.
For reasons that will become clear shortly, we now restrict our consideration to a special class of gauge fields $W_i^a$; those that are planar $i=1,2$ and pure gauges, i.e. have a vanishing field strength tensor
\begin{equation}\label{zero Gij}
G_{i,j}\equiv\frac{\partial W_i}{\partial x_j}-\frac{\partial W_j}{\partial x_i}-i\frac{g}{\hbar}[W_i,W_j]=0
\end{equation}
 Recall that under gauge transformations -a clear discussion of non-Abelian gauge transformations within the framework of the Hamiltonian formalism is given in \cite{shikakhwa12} -
 , the gauge field $\vc{W}$ transforms as:
\begin{equation}\label{gauge transformation}
\vc{W}^\prime   =   U\vc{W}U^{-1}+\frac{i\hbar}{g}U\vc{\nabla}U^{-1}
\end{equation}
with the unitary operator $U$ having the general form
\begin{equation}\label{U}
U=\exp(\frac{i}{2}\vc{\Lambda}(\vc{x})\cdot\vc{\sigma})
\end{equation}
where $\vc{\Lambda}(\vc{x})$ is an arbitrary vector field.
%The  eigenfunctions of the Hamiltonian  transform as
%$\phi_{n}\rightarrow\phi'_{n}=U\phi_{n}=\exp(\frac{i}{2}\vc{\Lambda}(\vc{x})\cdot\vc{\tau})\phi_{n}$.
It is an established result \cite{Weinberg} that when
$\vc{W}$ is a pure gauge, then it is always possible to find a gauge transformation that sets $\vc{W}$ to zero:
\begin{equation}\label{remove W}
\vc{W}\rightarrow\vc{W}^\prime   =   U\vc{W}U^{-1}+\frac{i\hbar}{g}U\vc{\nabla}U^{-1}=0
\end{equation}
  Thus making it possible to write down a general expression of $\vc{W}$ in terms of $U$ :
\begin{equation}\label{pure gauge W}
\vc{W} =\frac{i\hbar}{g}U^{-1}\vc{\nabla}U
\end{equation}
For such a gauge field, implementing the gauge transformation, Eq.(\ref{remove W}), removes $\vc{W}$ and takes the gauge-covariant Hamiltonian, Eq.(\ref{gauge covariant H }), to that of a free particle. This transformation is at the same time equivalent to implementing a unitary transformation on the Hamiltonian;
\begin{equation}\label{GC unitary transf}
UH_{GC}(\vc{W})U^{-1}= H_{GC}(\vc{W'}=0)= H_{free}'
\end{equation}
$H_{GC}$ and $H_{free}'$ are unitarily-equivalent and isospectral, and the corresponding eigenfunctions  are related as:
\begin{equation}\label{wf transformation}
\psi'_{n}=U\psi_{n}
\end{equation}
As for our  relevant non gauge-covariant Hamiltonian, Eq.(\ref{HNGC}), a gauge transformation in the sense of Eq.(\ref{gauge transformation}), indeed takes the Hamiltonian to that of the free particle Hamiltonian as well. The issue, however, is that here, in contrast to the gauge-covariant case,  this gauge transformation can \emph{not} be implemented through a unitary transformation on $H$,i.e.
\begin{equation}\label{non GC unitary transf}
UH(\vc{W})U^{-1}=\frac{\vc{p}^{2}}{2m}-\frac{g^{2}}{2m}\vc{W}\cdot\vc{W}\equiv H'\neq H(\vc{W'}=0)
\end{equation}
In other words, the unitary transformation in the non gauge-covariant case is \emph{not} a gauge transformation !. Note, however, that the following relation still holds:
\begin{equation}\label{unitariy equivalence NGC}
 H(\vc{W})=U^{-1}\left(\frac{\vc{p}^{2}}{2m}-\frac{g^{2}}{2m}\vc{W}\cdot\vc{W}\right)U\equiv U^{-1}H'U
\end{equation}
which implies that the eigenfunctions of $H(\vc{W})$ and $H'$ are related as:
\begin{equation}\label{psi' in terms of psi}
\psi_{n}=U^{-1}\psi'_{n}
\end{equation}
Note also that although $H'\equiv UH(\vc{W})U^{-1} $ is not the free particle Hamiltonian, it is spin-independent since the term quadratic in the gauge field $\vc{W}$ is proportional to the identity as we have shown before.

\section{PSH in a plane}
We introduce a special class of the pure  gauge fields, the Abelian but at the same time $2\times2$ matrix pure gauge fields.  If the vector function
$ \vc{\Lambda}(\vc{x})$ appearing in $U$, Eq.(\ref{U}), is of the general form
\begin{equation}\label{Abelian lambda}
\vc{\Lambda}(\vc{x})= \lambda(\vc{x})\hat{\vc{n}}
\end{equation}
with $\hat{\vc{n}}$ a \emph{constant } unit vector, then Eq.(\ref{pure gauge W}) gives us the following special  form of $\vc{W}$:
\begin{equation}\label{abelian W}
\vc{W}=-\frac{\hbar}{g}(\hat{\vc{n}}\cdot\vc{\sigma})\vc{\nabla}(\frac{\lambda(\vc{x})}{2})
\end{equation}
which is evidently Abelian:
\begin{equation}\label{CR's}
[W_i,W_j]=0
\end{equation}
The unitary operator $U$, Eq.(\ref{U}), that defines - and at the same time gauges away- this Abelian gauge field has a special form now:
\begin{equation}\label{abelian U}
U=\exp i\{(\frac{\lambda(\vc{x})}{2})(\hat{\vc{n}}\cdot\vc{\sigma})\}
\end{equation}
The above is just a rotation operator about the \emph{fixed} direction defined by $\hat{\vc{n}}$, but with the \emph{space-dependent} angle $\lambda(\vc{x})$.Eqs.(\ref{abelian W}) and (\ref{abelian U})are, respectively, the most general form and the corresponding unitary transformation for a planar, Abelian pure gauge field. These,with the simple relation between the rotation angle $\lambda(\vc{x})$ and the general form of the gauge field are reported for the first time to the best of our knowledge.   Note that it is this specific form of the gauge field that is behind the phenomenon of PSH as we will show below.\\
 Here, we first introduce the general algorithm for gauging away any Abelian gauge field of the form given above, and then apply this algorithm to the case of an electron gas subject to an equal-strength R + D SOI in which the PSH was discovered.
%We now demonstrate that the phenomenon of PSH arises for \emph{any} SOI that has  the general form of the Abelian field given by Eq.(\ref{abelian W}). We then apply the general formalism to the  experimentally realizable case, namely,a planar system  with both the R and D SOI's present and having equal strengths.
 To this end, if we read the gauge and unitary transformations, Eq.(\ref{gauge transformation}), and Eqs.(\ref{non GC unitary transf})-(\ref{psi' in terms of psi}), for the Abelian pure gauge, Eq.(\ref{abelian W}), we see that the eigenfunctions $\psi$ of the SOI Hamiltonian $H(\vc{W})$ are constructed from  those of the spin-independent Hamiltonian $H'$ by just multiplying them by the space-dependent rotation about $\hat{\vc{n}}$, Eq.(\ref{abelian U}), (recall that $ U^{-1}=U^\dagger = \exp -i\{(\frac{\lambda(\vc{x})}{2})(\hat{\vc{n}}\cdot\vc{\sigma})\}$). The space-dependence of this rotation is such that it guarantees that the spin projection about $\hat{\vc{n}}$ is conserved. If $H'$ has propagating solutions, then the spin direction of the particle will precess about $\hat{\vc{n}}$ as it propagates, with the angle of precession being $\lambda(\vc{x})$. It is the appearance of this phenomenon in a free electron gas that gave rise to the name PSH. Note that both the unit vector $\hat{\vc{n}}$ and the precession angle $\lambda(\vc{x})$ are fixed by the specific form of the gauge field , i.e. by the SOI. In the same manner, for non-propagating solutions of $H'$, in the corresponding solutions of $H(\vc{W})$ we will have the spin of the  particle rotated about $\hat{\vc{n}}$  with different angles according to the spacial distribution of the eigenfunctions.

We now apply the above formalism to the case when R + D SOI's with equal strengths are present. Let us recall that the  Hamiltonian for the RSOI is \cite{rashba},
\begin{equation}
  H^R=\frac{\vc{p}^{2}}{2m}+\frac{\alpha}{\hbar}\left(p_{y}\sigma_{x}-p_{x}\sigma_{y}\right)~~,
  \label{RSOI}
\end{equation}
and that for the DSOI coupling\cite{dresselhaus} is
\begin{equation}\label{DSOI}
  H^D=\frac{\vc{p}^{2}}{2m}+\frac{\beta}{\hbar}\left(p_{x}\sigma_{x}-p_{y}\sigma_{y}\right)
\end{equation}
here $\alpha$ and $\beta$ are the strengths of the SOI couplings and are taken as mere constants. The special case
when both the RSOI and DSOI are present and have equal strengths, i.e. $\alpha=\pm\beta$  will be our focus as
 it is the system where PSH phenomenon appears.  The Hamiltonian for this case can be cast in the form:
\begin{equation} \label{ H R+D}
  H^{R+D}_{\alpha=\pm\beta} =\frac{\vc{p}^{2}}{2m}
           +\frac{\alpha}{\hbar}\left(\sigma_{x}\mp\sigma_{y}\right)
           \left(p_{y}\pm p_{x}\right)
\end{equation}
To apply the formalism developed above to this case, we note that this Hamiltonian is of our standard form given by Eq.(\ref{standard H}) upon identifying the components of $\vc{W}$ as :
\begin{eqnarray}\label{Wx and Wy}
-\frac{g}{m}W_x|_{\alpha=\pm\beta}&=&\pm\frac{\alpha}{\hbar}\left(\sigma_{x}\mp\sigma_{y}\right)=\pm\frac{\alpha}{\hbar}(\hat{\vc{n}}_{\pm}\cdot\vc{\sigma})\\\nonumber
-\frac{g}{m}W_y|_{\alpha=\pm\beta}&=&\frac{\alpha}{\hbar}\left(\sigma_{x}\mp\sigma_{y}\right)=\frac{\alpha}{\hbar}(\hat{\vc{n}}_{\pm}\cdot\vc{\sigma})
\end{eqnarray}
with the unit vector  $\hat{\vc{n}}_{\pm}$ characteristic of the R+D SOI (at $\alpha=\pm\beta$) given  by:
\begin{equation}\label{n}
 \hat{\vc{n}}_{\pm}=\frac{\hat{ \vc{i}}\mp \hat{ \vc{j}}}{2}
\end{equation}
It is immediately evident that $[W_x,W_y]=0$ and since $\alpha$ is a constant, it  is a pure gauge. Putting it in the form given by Eq.(\ref{abelian W})we have
\begin{equation}\label{abelian W R+D}
\vc{W}^{\alpha=\pm\beta}=-\frac{\hbar}{g}(\hat{\vc{n}}\cdot\vc{\sigma})\left(\frac{m\alpha}{\hbar^2}\hat{ \vc{j}}\pm \frac{m\alpha}{\hbar^2}
\hat{ \vc{i}}\right)= -\frac{\hbar}{g}(\hat{\vc{n}}\cdot\vc{\sigma})\vc{\nabla}(\frac{\lambda(\vc{x})}{2})
\end{equation}
thus identifying
\begin{equation}\label{gradient lambda}
\vc{\nabla}(\frac{\lambda(\vc{x})}{2})|_{\alpha=\pm\beta}=\left(\frac{m\alpha}{\hbar^2}\right)(\pm\hat{ \vc{i}}+\hat{ \vc{j}})
\end{equation}
Here, we note that in this specific case, the gauge-symmetry breaking term in the Hamiltonian is just a constant since $\frac{g^{2}}{2m}\vc{W}\cdot\vc{W}=\frac{m\alpha^2}{\hbar^2}$.
Therefore, the eigenfunctions $\psi_n=U^{-1}\psi'_n$ of the Hamiltonian, Eq.(\ref { H R+D}), are just those of a spin one-half  free particle  multiplied by the position-dependent rotation $U$ about the fixed axis $\hat{\vc{n}}_{\pm}=\frac{\hat{ \vc{i}}\mp \hat{ \vc{j}}}{2}$. Now, what remains is to integrate to find $\lambda$ and then construct $U$ given in  Eq,(\ref{abelian U})that removes $\vc{W}$. With
\begin{equation}\label{lambda R+D}
\frac{\lambda(\vc{x})}{2}|_{\alpha=\pm\beta}=\frac{m\alpha}{\hbar^2}(y\pm x)
\end{equation}
we have
\begin{equation}\label{U R+D}
U_{R+D}=\exp i\{{\frac{m\alpha}{\hbar^2}(y\pm x)(\hat{\vc{n}}_{\pm}\cdot\vc{\sigma})}\}
\end{equation}
%with the space-dependent rotation angle being $ \lambda(\vc{x})=\frac{2m\alpha}{\hbar^2}(y\pm x)$
One can readily check that this $U$ indeed takes $\vc{W}$ given by Eq.(\ref{abelian W R+D}) to zero. The above equation has been derived  in reference \cite{chen&chang08}.
The value added in this derivation is that it is a special case of a general formalism of any linear SOI that is expressible as a pure gauge field.

\section{PSH on a ring}
The Planar Hamiltonian with SOI expressed as a gauge field in polar coordinates reads:
\begin{equation}\label{planar H}
H= \frac{-\hbar^2}{2m}
\left(\frac{1}{r}\frac{\partial}{\partial r}(r\frac{\partial}{\partial r })+\frac{1}{r^2}\frac{\partial^2}{\partial\theta^2}
\right)+i\frac{\hbar g}{m}\left(W_{r}\frac{\partial}{\partial r}+\frac{1}{r}W_{\theta}\frac{\partial}{\partial\theta}
\right)
%- \frac{g^{2}}{2m}\left(W_r^2+W^2_{\theta}\right)
\end{equation}
The construction of the Hermitian ring Hamiltonian is not a trivial issue. We will follow the approach introduced in \cite{shikakhwa pla1, shikakhwa pla2} (see also \cite{shikakhwa ejp})to do this. We start from the
above Hamiltonian and assume that a strong radial potential $V(r)$ -not shown in the Hamiltonian- confines it within a narrow layer about a ring of radius $a$, say. In the limit of a very strong potential the particle  is then pinned to the ring. As was shown in \cite{shikakhwa pla1, shikakhwa pla2} , the Hermitian radial momentum operator is this case is not simply $-i\hbar\frac{\partial}{\partial r}$, rather it is the operator
\begin{equation}\label{hermitian p}
p_r\equiv -i\hbar(\frac{\partial}{\partial r}+\frac{1}{2r})
\end{equation}
So, we can express the planar Hamiltonian, Eq.(\ref{planar H}), in terms of $p_r$ as:
\begin{equation}\label{}
H=\frac{(p_r)^2}{2m}-\frac{\hbar^2}{2m}\left(\frac{1}{r^2}\frac{\partial^2}{\partial\theta^2}\right)-\frac{\hbar^2}{8mr^2}-\frac{g}{m}W_rp_r-
\frac{i\hbar g}{m}\frac{W_r}{2r}+\frac{i\hbar g}{m}\left(\frac{1}{r}W_\theta\frac{\partial}{\partial\theta}\right)
\end{equation}
When the particle is pinned to the surface, the radial degree of freedom is frozen, and so we can drop the momentum operator $p_r$ from the above Hamiltonian. Setting also $r=a$ and substituting (for a constant $\vc{W}$ and  as a result of the condition $\vc{\nabla}\cdot\vc{W}=0$) $\frac{W_r}{2r}=-\frac{1}{2r}\frac{\partial}{\partial\theta}$ we have the Hermitian ring Hamiltonian as:
\begin{equation}\label{H ring NGC extended}
H^{ring}=-\frac{\hbar^2}{2ma^2}\frac{\partial^2}{\partial\theta^2}-\frac{\hbar^2}{8ma^2}+\frac{i\hbar g}{ma}W_\theta\frac{\partial}{\partial\theta}+\frac{i\hbar g}{2ma}\frac{\partial W_{\theta}}{\partial\theta }
\end{equation}
Note that while any reference to $W_r$ dropped from the Hamiltonian along with $p_r$ , this component is still there ! it is just not "seen" on the ring. In fact, it should always be there to preserve the condition $\vc{\nabla}\cdot\vc{W}=\frac{\partial W_{r}}{\partial r}+\frac{W_r}{2a}+\frac{1}{a}\frac{\partial W_{\theta}}{\partial\theta}=0$, which should be met everywhere in the plane including on the ring.
The above  is now the most general  Hamiltonian with a linear and constant  SOI  for  a spin one-half  particle on a ring. It can be also put in the form:
\begin{equation}\label{H ring NGC}
H^{ring}=-\frac{\hbar^2}{2ma}\left(\frac{1}{a}\frac{\partial}{\partial \theta}-\frac{ig}{\hbar}W_{\theta}\right)^2-\frac{\hbar^2}{8ma^2}-\frac{g^{2}}{2m}W_{\theta}^2
\end{equation}
The above, of course, is the non gauge-covariant Hamiltonian on the ring constructed starting from the non gauge-covariant planar Hamiltonian. $-\frac{g^{2}}{2m}W_{\theta}^2$ is the gauge-symmtry breaking term. For an Abelian $\vc{W}$ of the form given in Eq.(\ref{abelian W}), we have $W_r$ and $W_{\theta}$ as:
\begin{eqnarray}
  W_r &=& -\frac{\hbar}{g}(\hat{\vc{n}}\cdot\vc{\sigma})\hat{\vc{r}}\cdot\vc{\nabla}(\frac{\lambda(\vc{x})}{2})
=-\frac{\hbar}{g}(\hat{\vc{n}}\cdot\vc{\sigma})\frac{\partial }{\partial r}\frac{\lambda (\vc{r})}{2} \\
  W_{\theta} &=& -\frac{\hbar}{g}(\hat{\vc{n}}\cdot\vc{\sigma})\hat{\vc{\theta}}\cdot\vc{\nabla}(\frac{\lambda(\vc{x})}{2})
=-\frac{\hbar}{g}(\hat{\vc{n}}\cdot\vc{\sigma})\frac{1}{a}\frac{\partial }{\partial \theta}\frac{\lambda (\vc{r})}{2}
\end{eqnarray}
Similarly, the gauge transformation, Eq.(\ref{gauge transformation}) reads for the polar components of the above $\vc{W}$:
\begin{eqnarray}
  W'_r &=& W_r+\frac{i\hbar}{g}U\frac{\partial}{\partial r}U^{-1} \\\nonumber
   W'_{\theta} &=& W'_{\theta}+\frac{i\hbar}{g}U\frac{1}{a}\frac{\partial}{\partial \theta}U^{-1}
\end{eqnarray}

 For  $U$ in Eq.(\ref{abelian U}), which is  $U$ in the planar case,  the above gauge transformation sets \emph{both} polar components of $\vc{W}$ to zero. Applying this transformation to the ring Hamiltonian, Eq.(\ref{H ring NGC}), will take it , therefore, to $H^{ring}(\vc{W}=0)$ :
 \begin{equation}\label{H ring W=0}
H^{ring}(\vc{W}=0)=-\frac{\hbar^2}{2ma^2}\frac{\partial^2}{\partial\theta^2}-\frac{\hbar^2}{8ma^2}
 \end{equation}
 which is - up to a constant- that of a "free"  particle on a ring. In this case, too, as  in the planar case ( see Eq.(\ref{non GC unitary transf}) and the discussion below it) this gauge transformation is NOT equivalent to a unitary transformation on the Hamiltonian. This is because:
 \begin{equation}\label{non GC unitary trans ring}
  UH^{ring}U^{-1}=-\frac{\hbar^2}{2ma^2}\frac{\partial^2}{\partial\theta^2}-\frac{\hbar^2}{8ma^2}
  -\frac{g^{2}}{2m}W_{\theta}^2\neq H^{ring}(\vc{W}=0)
 \end{equation}
 Note that even in the case when $\vc{W}$ is a constant Cartesian vector, $W_{\theta}$ need not and generally is not a constant; it is a function of $\theta$. The above, therefore, is the Hamiltonian of a particle on the ring subject to a $\theta$-dependent potential.
 %Depending on the specific form of $W_\theta$ it may develop bound states.
 Regardless of the absence of gauge-covariance, we still have the following transformation holding:
 \begin{equation}\label{}
H^{ring}= U^{-1}\left(H^{ring}(\vc{W}=0)-\frac{g^{2}}{2m}W_{\theta}^2\right)U\equiv U^{-1}H_{ring}'U
 \end{equation}
So, the eigen functions $\psi_n$ of the SOI Hamiltonian on the ring  $H^{ring}$ are related to those of $H_{ring}'$ defined in the above equation  as $\psi_n=U^{-1}\psi'_n$ with U again the same operator defined in Eq.(\ref{abelian U}). This means that they are related by a position-dependent rotation about the fixed axis $\hat{\vc{n}}$ , thus  the spin projection along this direction is still conserved.  It is important to note her that we should gauge away $W_r$ too along with $W_\theta$, even though it is not "seen" in the ring Hamiltonian. This is because the condition $\vc{\nabla}\cdot\vc{W}=\vc{\nabla}\wedge\vc{E}=0$ will be violated otherwise.

It is now straightforward to apply the above formalism to the R+D SOI when $\alpha=\pm\beta$. The gauge field components in polar coordinates are now given as:
\begin{eqnarray}\label{Wr and w theta}
  W_r &=& -\frac{\hbar}{g}(\hat{\vc{n}}\cdot\vc{\sigma})\frac{\partial }{\partial r}\frac{\lambda (\vc{r})}{2}= -\frac{\hbar}{g} (\hat{\vc{n}}_{\pm}\cdot\vc{\sigma})\left(\frac{m\alpha}{\hbar^2} \right)(\sin\theta\pm\cos\theta) \\\nonumber
  W_\theta &=& -\frac{\hbar}{g}(\hat{\vc{n}}\cdot\vc{\sigma})\frac{1}{a}\frac{\partial }{\partial \theta}\frac{\lambda (\vc{r})}{2}=-\frac{\hbar}{g} (\hat{\vc{n}}_{\pm}\cdot\vc{\sigma})\left(\frac{m\alpha}{\hbar^2} \right)(\cos\theta\mp\sin\theta)
\end{eqnarray}
The ring Hamiltonian, upon substituting the above expressions in Eq.(\ref{H ring NGC}) will thus read:
\begin{equation}\label{H ring R+D}
H^{ring}_{R+D}=-\frac{\hbar^2}{2m}\left(\frac{1}{a}\frac{\partial}{\partial \theta}+\frac{im\alpha}{\hbar^2}(\hat{\vc{n}}_{\pm}\cdot\vc{\sigma})(\cos\theta\mp\sin\theta)\right)^2
-\frac{\hbar^2}{8ma^2}-\frac{m\alpha^{2}}{2\hbar^2}(1\mp\sin2\theta)
\end{equation}
The explicit form of U, Eq.(\ref{abelian U}), is straightforward to calculate and is given by:
\begin{equation}\label{U ring R+D}
U_{R+D}^{ring}=\exp i\{(\frac {m\alpha}{\hbar^2})(\hat{\vc{n}}_{\pm}\cdot\vc{\sigma})a(\sin\theta\pm\cos\theta)\}
\end{equation}
which is just -as expected- the polar coordinate form of $U$ in the plane found earlier, Eq.(\ref{U R+D}), evaluated at $r=a$ . This transformation, when applied to the Hamiltonian, Eq.(\ref{H ring R+D}), gives :
\begin{equation}\label{H' R+D ring}
H'\: ^{ring}_{R+D}\equiv U^{-1}H^{ring}_{R+D}U=-\frac{\hbar^2}{2ma^2}\frac{\partial^2}{\partial \theta^2}
-\frac{\hbar^2}{8ma^2}-\frac{m\alpha^{2}}{2\hbar^2}(1\mp\sin2\theta)
\end{equation}
This Hamiltonian was obtained in \cite {sheng&chang06}without using the gauge field formalism, however. The solutions are generally the Mathieu functions \cite{Arfken}. The eigenfunctions $\psi_{ring}$ of the original SOI ring Hamiltonian,Eq.(\ref{H ring R+D}), can be constructed from those of the above $H'\: ^{ring}_{R+D}$ as :
\begin{equation}\label{psi ring from psi' }
\psi^{ring}_{R+D}=(U_{R+D}^{ring})^{-1}\psi '\:^{ring}_{R+D}
\end{equation}

\section{Conclusions}
In this work, we have considered the question of  systematically formulating the theory  of SOI of a spin one-half particle in termsof a pure ( thus can be gauged out) non-Abelian gauge field $\vc{W}$. The corresponding Hamiltonian is not gauge-covariant due to the presence of a term $\sim\vc{W}\cdot\vc{W}$ that breaks the gauge symmetry. As a consequence, we have shown that the gauge transformation that gauges it out is not equivalent to a unitary transformation of the Hamiltonian as is the case with a gauge-covariant Hamiltonian. This latter unitary transformation  removed $\vc{W}$ only partially from the Hamiltonian.  Nevertheless, the eigenfunctions of the two unitarily-equivalent Hamiltonians  are related by this unitary transformation that partially removes $\vc{W}$ ( see Eqs.(\ref{GC unitary transf})-(\ref{psi' in terms of psi})).\\
A general form (Eq.(\ref{abelian W})) was constructed for a sub-class of these gauge fields that , though $2\times2$ matrices, are Abelian. It was shown that the corresponding unitary transformation that partially  removed these gauge fields from the Hamiltonian had the form of a rotation about a fixed axis but with a space-dependent angle, both of which being characteristics of the gauge field, and thus the specific SOI present. The case of a spin one-half particle subject to R+D SOI was shown to be a special case of this class of gauge fields, and the phenomenon of PSH that appears in these systems was seen to be a natural consequence of the general formalism introduced.  The fact that the gauge symmetry-breaking term was in this case a constant was noted to be behind the PSH phenomenon. \\

We have then carried over the  formalism to the case of a particle confined to a ring. It has been  shown that upon constructing the Hamiltonian for the particle on the ring, all reference to the radial component of the gauge field dropped  and only the angular component survived. It was stressed, however, that the radial component was still there, though not seen on the ring itself. This was crucial to preserve the Coulomb gauge condition $\vc{\nabla}\cdot\vc{W}=0$, it was argued. Again, in this case also, the unitary transformation that partially removed the $\theta$-component of the gauge field was shown to be the same as that in the planar case, and it also simultaneously removed  the 'unseen" radial  component of the gauge field; again to preserve the gauge condition. The general formalism was again applied to the the case of equal strength R+D SOI, and the Hamiltonian of a spin one-half particle subject to an angular potential reported in the literature - not using the gauge field approach- was obtained . The Unitary transformation that related the eigenfunctions of the two Hamiltonians was seen again to be the same as the one in the corresponding planar case.

% The gauge-covariant case is a bit more subtle. To appreciate this, note that the planar Hamiltonian to start with in this case is
%\begin{equation}\label{planar GC H}
%H_{NGC}=\frac{\left(\vc{p}-g\vc{W}\right)^{2}}{2m}=\frac{-\hbar^2}{2m}
%\left(\frac{1}{r}\frac{\partial}{\partial r}(r\frac{\partial}{\partial r })+\frac{1}{r^2}\frac{\partial^2}{\partial\theta^2}
%\right)+i\frac{\hbar g}{m}\left(W_{r}\frac{\partial}{\partial r}+\frac{1}{r}W_{\theta}\frac{\partial}{\partial\theta}
%\right)- \frac{g^{2}}{2m}\left(W_r^2+W^2_{\theta}\right)
%\end{equation}
%Therefore, upon applying the same confinement procedure to a ring as above, we will have an extra $\frac{g^{2}}{2m}W_r^2$ term as this last does not drop from H along with $p_r$ as in the non gauge-covariant case;
%\begin{eqnarray}
% H^{ring}_{GC} &=& -\frac{\hbar^2}{2ma^2}\frac{\partial^2}{\partial\theta^2}-\frac{\hbar^2}{8ma^2}+\frac{i\hbar g}{ma}W_\theta\frac{\partial}{\partial\theta}+\frac{i\hbar g}{2ma}\frac{\partial W_{\theta}}{\partial\theta}-\frac{g^{2}}{2m}W_r^2 \\\nonumber
%  &=& -\frac{\hbar^2}{2ma}\left(\frac{1}{a}\frac{\partial}{\partial \theta}-\frac{ig}{\hbar}\right)^2-\frac{\hbar^2}{8ma^2}-
 % \frac{g^{2}}{2m}\left(W_r^2+W_{\theta}^2\right)
%\end{eqnarray}\label{H ring GC}

\end{document}